\documentstyle[12pt,equation]{article}
\begin{document}
\def\ub{\underbar}
\def\wgg{\mbox{$W_{\gamma\gamma}$}}
\def\be{\begin{equation}}
\def\ee{\end{equation}}
\newcommand{\een}{\end{subequations}}
\newcommand{\ben}{\begin{subequations}}
\newcommand{\beq}{\begin{eqalignno}}
\newcommand{\eeq}{\end{eqalignno}}
\def\bea{\begin{eqnarray}}
\def\eea{\end{eqnarray}}
\def\psq{\mbox{$P^2$}}
\def\qsq{\mbox{$Q^2$}}
\def\lsq{\mbox{$\Lambda^2$}}
\def\xg{\mbox{$x_{\gamma}$}}
\def\gamg{${ \gamma g }$}
\def\gamgam{\mbox{$\gamma \gamma $}}
\def\gamp{\mbox{$\gamma p$}}
\def\eplem{\mbox{$e^+ e^- $}}
\def\as{\mbox{$ \alpha_s$}}
\def\rts{\mbox{$ \sqrt{s} $}}
\def\fGg{\mbox{$f_{G|\gamma}$}}
\def\fig{\mbox{$f_{i|\gamma}$}}
\def\fqg{\mbox{$f_{q|\gamma}$}}
\def\fie{\mbox{$f_{i|e}$}}
\def\fGe{\mbox{$f_{G|e}$}}
\def\fqe{\mbox{$f_{q|e}$}}
\def\qg{\mbox{$q^{\gamma}$}}
\def\Gg{\mbox{$G^{\gamma}$}}
\def\xqsq{\mbox{$(x,Q^2)$}}
\def\pc{\mbox{$P_c$}}
\def\pcsq{\mbox{$P_c^2$}}
\def\pminsq{\mbox{$P^2_{\rm min}$}}
\def\pmaxsq{\mbox{$P^2_{\rm max}$}}
\def\fge{\mbox{$ f_{\gamma/e}$}}

\renewcommand{\thefootnote}{\fnsymbol{footnote}}

\pagestyle{empty}
\begin{flushright}
MAD/PH/819 \\
BU 94--02 \\
March 1994\\
\end{flushright}
\vspace*{1cm}
\begin{center}
{\Large \bf Virtual Photon Structure Functions and the Parton Content of the
Electron}\\
\vspace*{5mm}
Manuel Drees$^1$\footnote{Heisenberg fellow} and Rohini M. Godbole$^2$ \\
\vspace*{5mm}
$^1${\it Physics Dept., Univ. of Wisconsin, 1150 University Ave.,
Madison, WI 53706, USA}\\
\vspace*{5mm}
$^2${\it Physics Dept., Bombay Univ., Vidyanagari, Santa Cruz (East),
Bombay 400098, India} \\
\end{center}
\begin{abstract}
We point out that in processes involving the parton content of the photon the
usual effective photon approximation should be modified. The reason is that
the parton content of virtual photons is logarithmically suppressed compared
to real photons. We describe this suppression using several simple, physically
motivated ans\"atze. Although the parton content of the electron in general
no longer factorizes into an electron flux function and a photon structure
function, it can still be expressed as a single integral. Numerical examples
are given for the \eplem\ collider TRISTAN as well as the $ep$ collider
HERA.
\end{abstract}
\clearpage
\pagestyle{plain}
\setcounter{page}{1}

\section*{1) Introduction}
Resolved photon processes \cite{14} are now being studied in some detail at
both \eplem\ colliders (TRISTAN \cite{1,2}, LEP \cite{3}) and the $ep$
collider HERA \cite{4,5}. These are processes involving the quark and gluon
``content" of the photon \cite{6}. The immediate goal of studying such
reactions is to determine the photon structure functions experimentally, i.e.
to test which (if any) of the parametrizations that have been proposed [8--12]
reproduce the data. Ultimately one hopes to gain new insight into QCD
\cite{6,12} from such studies. A somewhat more mundane but still quite
important task is to reduce uncertainties, due to our lack of knowledge of the
parton content of the photon, in predictions of hadronic backgrounds at future
high--energy \eplem\ \cite{13,13a} and $ep$ colliders.

All existing theoretical estimates \cite{14} of resolved photon cross sections
make use of the Weizs\"acker--Williams or effective photon approximation
\cite{15} to translate \gamgam\ and \gamp\ cross sections into \eplem\ and
$ep$ cross sections. The same formalism has been used when data [2--6] have
been compared to theoretical expectations. Assuming that experimental
(anti--)tagging requirements as well as non--logarithmic terms in the photon
flux function are properly taken into account this approximation has been
shown \cite{16,17,23} to reproduce quite accurately exact calculations of
processes where the photon participates directly, i.e. is not resolved into
its hadronic substructure. However, no such check exists for resolved photon
processes. Such a check would necessitate a complete understanding of the
dependence of the parton content of the photon on the photon's virtuality
\psq. While this dependence is computable \cite{18,19} from perturbative QCD
for large $\psq \gg \lsq$, and can be assumed to be negligible for $\psq \ll
\lsq$, no satisfying treatment for the transition region $\psq \sim \lsq$
exists. On the other hand, since contributions from far off--shell photons are
suppressed by the photon propagator $1/\psq$, the contribution from $\psq \sim
\lsq$ (or less) is usually numerically more important than the theoretically
clean high--\psq\ region.

In ref.\cite{13} we gave a first crude estimate of the suppression due to the
virtuality of the exchanged photon. Here we attempt a more careful treatment,
making use of recent results by Borzumati and Schuler \cite{19}, who pointed
out that quark and gluon densities should be treated separately, the
suppression being more severe in the latter case. Since the region of
intermediate \psq\ cannot (yet) be treated rigorously we use several simple
ans\"atze that contain one free parameter and reproduce the correct
high--\psq\ limit. We compare these with parameter--free predictions based on
simple quark--parton model (QPM) calculations, and find reasonable agreement.

The remainder of this article is organized as follows. In sec.2 we describe
the general framework, taking care to treat experimental (anti--)tagging of
outgoing electrons properly. In secs. 3 and 4 this formalism is applied to
quark and gluon densities, respectively. In all cases we were able to express
the experimentally relevant quantity, the parton density (flux) function ``in"
the electron, in terms of a single integral to be computed numerically; the
resulting expressions for cross sections are then as readily treatable as
existing ones that ignore the virtuality of the exchanged photons. In sec.5
some numerical results are presented. Not surprisingly, anti--tagging,
which imposes an upper limit on \psq, reduces the suppression, and it vanishes
altogether if a forward tagger is used (as is done by the HERA experiments). A
recently installed small angle electron detector should be able to study
virtual photon effects in some detail, when combined with the existing forward
tagger. Finally, sec.6 contains a brief summary and some conclusions.

\section*{2) General formalism}
Since quite detailed discussions of the Weizs\"acker--Williams approximation
already exist in the literature \cite{20,21,16,22} we can be brief in this
section. Consider a reaction that proceeds via the exchange of a photon in
the $t-$ or $u-$ channel, $e+X \rightarrow e + X'$. In the effective photon
approximation the corresponding cross section is then written as: \be
\label{e1}
d \sigma (eX \rightarrow e X') \simeq \fge(\xg) d\xg \cdot d \sigma(\gamma X
\rightarrow X'), \ee
where $\xg \equiv E_{\gamma}/E_e$ is the scaled photon energy. This
approximation is valid if \\
{\it (i)} The contribution from the exchange of longitudinal photons is
negligible; and \\
{\it (ii)} For the bulk of the contribution to the exact cross section the
photon virtuality \psq\ is small compared to the scale \qsq\ characterizing
the process $\gamma + X \rightarrow X'$.\\
These conditions are necessary since eq.(\ref{e1}) expresses the cross section
for $eX$ scattering in terms of a cross section for $\gamma X$ scattering
where the photon is {\em on--shell}, i.e. purely transverse.

If these conditions are fulfilled the photon flux can be written as
\cite{22,23} \be \label{e2}
\fge(\xg) = \int_{P^2_{\rm min}}^{P^2_{\rm max}} \frac {dP^2} {P^2}
\tilde{f}(\xg) - \frac {\alpha}{\pi} m_e^2 \xg \left( \frac {1}{\pminsq} -
\frac {1} {\pmaxsq} \right), \ee
where $m_e$ is the electron mass and \be \label{e3}
\tilde{f}(\xg) = \frac {\alpha} {2\pi x_{\gamma}}
\left[ 1 + (1-\xg)^2 \right]. \ee
The kinematical limits on the virtuality are \ben \label{e4} \beq
P^2_{\rm{ min, kin}} &= m_e^2 \frac {x^2_{\gamma}} {1-x_{\gamma}} ;
\label{e4a} \\
P^2_{\rm{max, kin}} &= 0.5 s \cdot (1-\xg) \cdot (1-\cos \! \theta_{\rm max}),
\label{e4b} \eeq \een
where $s$ is the squared centre--of--mass energy of the $eX$ system. In
eq.(\ref{e4b}) we have allowed for anti--tagging by introducing a maximal
scattering angle $\theta_{\rm max}$ of the outgoing electron (in the $eX$
cms frame). Similarly, small--angle tagging might introduce a lower bound on
the virtuality, $P^2_{\rm{min, tag}}$ that supersedes (\ref{e4a}). Moreover,
condition {\it (ii)} implies that the ansatz (\ref{e1}) breaks down if $\psq
> \qsq$; this has been confirmed in studies \cite{17,23} where
eqs.(\ref{e1})--(\ref{e3}) were compared to exact calculations of
$d \sigma ( eX \rightarrow e X')$. Altogether one thus has: \be \label{e5}
\fge(\xg) = \tilde{f}(\xg) \ln \frac {\pmaxsq} {\pminsq} + f_{\rm rest}(\xg),
\ee
with \ben \label{e6} \beq
\pminsq &= \max(P^2_{\rm{min, kin}}, P^2_{\rm{min, tag}}) ; \label{e6a} \\
\pmaxsq &= \min(P^2_{\rm{max, kin}}, \qsq) ; \label{e6b} \\
f_{\rm rest}(\xg) &= - \frac{\alpha}{\pi} \frac{1-x_{\gamma}}{x_{\gamma}}
\hspace*{3cm} {\rm if} \ P^2_{\rm{min, tag}} = 0; \nonumber \\
{} &= 0 \hspace*{4.4cm} {\rm if} \ P^2_{\rm{min, tag}} \gg m_e^2.
\label{e6c} \eeq \een
In writing eq.(\ref{e6c}) we have assumed $\pmaxsq \gg m_e^2$, which is true
for all applications at present high energy experiments.

The result (\ref{e5}) has been derived from eq.(\ref{e2}) under the assumption
that the only relevant \psq\ dependence is contained in the explicit factor
$1/\psq$. The standard procedure for treating resolved photon interactions
\cite{14} is to use eq.(\ref{e5}) to define a parton density inside the
electron: \be \label{e7}
f_{i|e}(y,\qsq) = \int_y^1 \frac {dx}{x} \fge \left( \frac{y}{x} \right)
f_{i|\gamma} \xqsq, \ee
where \fig \xqsq\ is the probability to find parton $i$ with momentum
fractrion $x$ in a {\em real} photon when probed at scale \qsq. The
cross section is then \be \label{e8}
d \sigma_{\rm res} (eX \rightarrow eX') = \sum_{i=q,G} f_{i|e} \xqsq
dx \cdot d \sigma(iX \rightarrow X'). \ee

The point we wish to make in this paper is that in case of resolved photon
interactions there is additional \psq\ dependence beyond the $1/\psq$ factor
contained in eq.(\ref{e2}). Of course, the exact cross section (\ref{e1}) will
always contain additional \psq\ dependence; however, in many cases this
dependence appears as terms $\propto \left( \psq / \qsq \right)^{n>0}$, which
can be neglected if condition {\it (ii)} is satisfied. The crucial difference
in case of resolved photon interactions is that they introduce an additional
(hadronic) scale, very roughly characterized by the the QCD scale parameter
$\Lambda$. This opens the possibility that terms $\propto \left( \psq / \lsq
\right)^n$ appear, which are {\em not} always small even if {\it (ii)} is
fulfilled. As we will see below, there is good reason to believe that the
leading \psq\ dependence is logarithmic; in other words, when writing
eq.(\ref{e7}) one ignores terms $\propto \ln (\psq / \lsq)$, which may not be
negligible compared to the leading terms $\propto \ln (\qsq / \lsq)$.

Fortunately we need not give up the effective photon approximation altogether,
since terms of the form $\ln (\psq / \lsq)$ can {\em only} originate from the
\psq\ dependence of the parton densities $f_{i|\gamma}$. We can therefore
generalize eqs.(\ref{e5}) and (\ref{e7}) in a straightforward manner:
\be \label{e9}
f_{i|e}(y,\qsq) = \int_y^1 \frac{dx}{x} \left[
\tilde{f} \left( \frac{y}{x} \right) \int_{P_{\rm min}^2}^{P_{\rm max}^2}
\frac {dP^2} {P^2} f_{i|\gamma} (x,Q^2,P^2 ) +
f_{\rm rest} \left(\frac{y}{x} \right) f_{i|\gamma} (x,Q^2,0) \right], \ee
where we have made use of the fact that $f_{\rm rest}$ is non--negligible only
if $\pminsq \ll \lsq$ so that real photon structure functions can be used in
the second term in eq.(\ref{e9}). This second term is thus the same as in
eqs.(\ref{e5}) and (\ref{e7}); for simplicity we will omit it from our
subsequent expressions, although it will be included in our numerical results.

The first term in eq.(\ref{e9}) involves a double integral, as opposed to the
single integral in the standard form (\ref{e7}). In order to make further
progress we must make some assumption regarding the \psq\ dependence of the
$f_{i|\gamma}$. As emphasized in ref.\cite{19} this dependence is quite
different for quarks and gluons; in the next two sections we therefore
discuss these two cases separately.

\section*{3) The quark density in the electron}
As mentioned above, the functions $f_{i|\gamma}(x,Q^2,P^2)$ can be computed
unambiguously from perturbative QCD in the kinematic region $\qsq \gg \psq
\gg \lsq$. Since a detailed literature on this topic already exists
\cite{18,19} we do not repeat this calculation here. The result is that the
parton densities are suppressed at high \psq\ compared to the case of real
photons; this is not surprising since a nonvanishing virtuality of the
photon implies a lower limit for the virtuality of the partons in that
photon. Unfortunately these rigorous, perturbative results are not applicable
in the region $\psq \simeq \lsq$. As discussed in the Introduction we expect
the contribution from this intermediate region to the inner integral in
eq.(\ref{e9}) to be at least as important numerically as the contribution from
the high--\psq\ region, due to the factor of $1/\psq$. Rather than attempting
to accurately reproduce the ($x-$dependent) suppression at large \psq\ as
predicted \cite{18,19} by QCD, we therefore use simple ans\"atze which
interpolate between the regions of low and high \psq.

We were guided by the observation of Borzumati and Schuler \cite{19} that the
parton densities inside a virtual photon approach the value predicted by the
simple QPM in the limit $\psq \rightarrow \qsq$, while the \psq\ dependence
disappears for $\psq \ll \lsq$. The simplest ansatz that incorporates this
behaviour is \beq \label{e10}
f^{(1)}_{q|\gamma}(x,Q^2,P^2) &= \qg\xqsq, \ \ \ \ \ \ \ \ \
\psq \leq \pcsq \nonumber \\
{ } &= c_q \xqsq \ln \frac {Q^2}{P^2}, \ \ \psq \geq \pcsq, \eeq
where \qg\xqsq\ are the standard quark density functions in real photons
\cite{6}. Continuity of the ansatz (\ref{e10}) at $\psq = \pcsq$ implies
\be \label{e11}
c_q \xqsq = \frac {\qg \xqsq} {\ln (Q^2 / P_c^2)}. \ee

In eq.(\ref{e10}) we are trying to describe the intricacies of nonperturbative
QCD in terms of a single parameter \pc. Clearly this cannot reproduce the
exact \psq\ dependence very accurately. However, here we are only interested
in the integral over \psq\ contained in eq.(\ref{e9}). Given that our ansatz
(\ref{e10}), as well as other ans\"atze to be described below, show the
correct limiting behaviour predicted by QCD it seems reasonable to believe
that these \psq\ integrals will indeed be described more or less correctly, if
\pc\ is chosen to be a typical hadronic scale, i.e. between a few hundred
MeV and a GeV.

Unlike the authors of ref.\cite{24} we do not distinguish between ``soft" and
``hard" components of the photon structure functions, where the soft
component (to be estimated from the Vector Dominance Model, VDM) would be
suppressed by a power of \psq\ (rather than logarithmically) at high \psq.
In this picture one assumes that the hard component is zero at some rather
low input scale $Q_0^2$, i.e. that at this scale the photon is indeed
identical to a vector meson as far as its hadronic properties are concerned.
It is not clear to us whether this amalgam of the VDM and QCD is indeed
meaningful. In any case, such a soft component could easily be incorporated in
our framework. As shown in ref.\cite{24}, for this soft part itself virtual
photon effects can to good approximation be included by simply cutting off the
\psq\ integral in eq.(\ref{e9}) at some scale $\simeq m^2_{\rho} \simeq 0.5
$ GeV$^2$. Our subsequent results would then only be valid for the hard part
of the photon structure functions, which can be obtained by subtracting
pion--like parton densities from standard parametrizations [8--12] of
\qg.

Inserting eqs.(\ref{e10}) and (\ref{e11}) into eq.(\ref{e9}) gives (recall
that we omit the term $\propto f_{\rm rest}$ here): \clearpage
\beq \label{e12}
f_{q|e}^{(1)}(y) = \int_y^1 \frac {dx}{x} \tilde{f} \left( \frac{y}{x} \right)
\qg \xqsq
&\left[ \ln \frac { \min(P_c^2,P^2_{\rm max}) } {P^2_{\rm min}}
\nonumber \right. \\ & \left.
+ \frac{1}{2} \theta(P_{\rm max}^2-P_c^2) \left( \ln \frac {Q^2}
{P_c^2} - \frac {\ln^2 (Q^2/ P^2_{\rm max})} {\ln (Q^2 / P_c^2)} \right)
\right], \eeq
where $\tilde{f}$ has been defined in eq.(\ref{e3}). This expression is
completely general; in particular, it allows to take (anti--)tagging into
account via its effects on \pminsq\ and \pmaxsq, see eqs.(\ref{e6}a,b). In the
important special case where there is no anti--tagging, i.e. where
$\pmaxsq=\qsq$, eq.(\ref{e12}) simplifies to: \be \label{e13}
f_{q|e}^{(1),{\rm no-tag}}(y) = \int_y^1 \frac {dx}{x} \tilde{f} \left(
\frac{y}{x} \right) \qg(x,Q^2) \ln \frac{Q^2} {P^2_{\rm min}} \left[ 1 -
\frac{1}{2} \frac { \ln  (Q^2 / P_c^2) } {\ln (Q^2 / P^2_{\rm min}) }
\right]. \ee
Notice that the integrands in eqs.(\ref{e12}) and (\ref{e13}) factorize into
a parton density (a function of $x$) and a photon flux factor (a function of
$x/y$) only if \pc\ is a constant, independent of $x$ (recall that \pminsq\
depends on the scaled photon energy $x/y$, see eq.(\ref{e4a})). However, even
in the general case where this factorization is lost \fqe\ is still given
by a single integral, just like in eq.(\ref{e7}) where virtual photon effects
have been ignored.

Eqs.(\ref{e10})--(\ref{e12}) can be used by simply assuming a constant value
for \pc; since it characterizes a typical hadronic scale, it should roughly
lie in the range \be \label{e14}
\lsq \leq \pcsq \leq m_p^2. \ee
Alternatively, one might try to estimate \pc\ from the QPM; after all, the
ansatz (\ref{e10}) was motivated by the QPM. One has (for quark mass
$m_q^2 \ll \psq$): \be \label{e15}
f^{\rm QPM}_{q|\gamma}(x,Q^2,P^2) = 3 \frac{\alpha}{2 \pi} e_q^2
\left[ x^2 + (1-x)^2 \right] \ln \frac {Q^2}{P^2} \equiv c_q^{\rm QPM}
\ln \frac {Q^2}{P^2}, \ee
which has the form of eq.(\ref{e10}). The QPM therefore makes a prediction
for $c_q$, so that eq.(\ref{e11}) can be solved for $\ln (Q^2/P_c^2)$; the
solution will in general depend on $x$ and \qsq.

The main advantage of the ansatz (\ref{e10}) is its simplicity. However,
when plotted vs. $\ln \psq$ it shows an ugly kink, i.e. the derivative
$\partial \fqg / \partial \ln \psq$ is discontinuous at $\psq=\pcsq$. This
drawback can be overcome by writing \be \label{e16}
f^{(2)}_{q|\gamma}(x,Q^2,P^2) = \qg \xqsq \frac {\ln \frac{Q^2+P_c^2}
{P^2+P_c^2}} {\ln \left( 1 + Q^2/P_c^2 \right) }. \ee
This modified ansatz has the same behaviour as eq.(\ref{e10}) in the
limits $\psq \rightarrow 0$ and $\psq \rightarrow \qsq$, but smoothely
interpolates between these two limits at $\psq \simeq \pcsq$. Strictly
speaking eq.(\ref{e16}) does not allow to express the \psq\ integral in
eq.(\ref{e9}) in terms of elementary functions. However, one can derive an
excellent analytical approximation to the exact result by splitting the
\psq\ integration into the domains $\psq \leq \pcsq$ and $\psq > \pcsq$,
using two different expansions for $\ln(\psq+\pcsq)$ in these integration
regions. The result is (for $\pmaxsq > \pcsq$): \beq \label{e17}
f_{q|e}^{(2)}(y) = \int_y^1 \frac {dx}{x} \tilde{f} \left( \frac {y}{x} \right)
\qg \xqsq & \left[ \ln \frac {P_c^2} {P^2_{\rm min}} + \frac{1}{2}
\ln \left( 1 + \frac {Q^2}{P_c^2} \right) \right. \nonumber\\ & \left.
+ \frac { P_c^2/P^2_{\rm max} + P^2_{\rm min}/P_c^2 - \pi^2/6 - 0.5 \ln^2
\frac {Q^2+P_c^2} {P^2_{\rm max}} } { \ln (1 + Q^2 / P_c^2)} \right] . \eeq
This result is exact up to terms ${\cal O} \left( \frac {P^4_c} {P^4_{\rm
max}}, \frac {P^4_{\rm min}} {P_c^4} \right)$; it reproduces the numerical
result to better than 2\% for all cases we tried.

Before presenting numerical predictions, we now turn to a discussion of gluon
densities.

\section*{4) The gluon density in the electron}
Unfortunately the simple ansatz (\ref{e10}) will not do for the case of
gluons. The reason is that, as emphasized in ref.\cite{19}, $\fGg(x,Q^2,P^2)$
vanishes faster than $\ln ( Q^2/P^2)$ as $P^2 \rightarrow Q^2$. This can be
understood perturbatively from the observation that a gluon has to be
radiated from a quark which is itself off--shell if $P^2 \ne 0$. One should
thus be able to find a reasonable ansatz for $\fGg(x,Q^2,P^2)$ by considering
a diagram where a photon splits into a $q \bar{q}$ pair and one of the quarks
radiates a gluon. Let $q_1^2$ and $q_2^2$ be the virtualities of the emitting
(anti--)quark and gluon, respectively; in the spirit of the backward showering
algorithm \cite{25} this gives for the gluon density: \be \label{e18}
\fGg(x,Q^2,P^2) \propto \int_{P^2}^{Q^2} \frac {d q_1^2}{q_1^2}
\int_{q_1^2}^{Q^2} \frac {dq_2^2}{q_2^2} \as. \ee

The result will obviously depend on the choice of momentum scale in \as,
which is ambiguous within the leading logarithmic approach followed here.
However, we know that the gluon density must vanish at least $\propto \ln^2
(Q^2/P^2)$ as $\psq \rightarrow \qsq$; on the other hand, for $\qsq \gg \psq$
we want to reproduce the well--known result \cite{6} that \fGg\ grows like
$\ln \! \qsq$. Chosing \as\ in eq.(\ref{e18}) to be independent of $q_1^2$
and $q_2^2$ gives $\fGg \propto \as \ln^2 (\qsq / \psq)$, which has the
correct high--\qsq\ behaviour only if we take the scale in \as\ to be \qsq.
This motivates the ansatz \beq \label{e19}
f^{(1a)}_{G|\gamma} (x,Q^2,P^2) &=
\Gg(x,Q^2), \hspace*{2.1cm} P^2 \leq P_c^2  \nonumber \\
&= c_G(x,Q^2) \frac {\ln^2 (Q^2 / P^2)} {\ln (Q^2 / \Lambda^2)},
\ P^2 \geq P_c^2, \eeq
where \Gg\xqsq\ is the gluon distribution function for on--shell photons.
Continuity at $\psq=\pcsq$ requires that \be \label{e20}
c_G\xqsq = \Gg \xqsq \frac {\ln ( Q^2 / \Lambda^2)} {\ln^2 ( Q^2 / P_c^2)},
\ee
which can easily be solved for \pcsq\ if $c_G$ is known (see below).

One obtains a slightly more complicated ansatz if \as\ in eq.(\ref{e18}) is
taken to depend on $q_1^2$ or $q_2^2$. Chosing $q_1^2$ as scale of \as\
leads to a result that grows faster than $\ln \! \qsq$ for large \qsq, which
is not acceptable. Taking $q_2^2$ as scale does lead to a reasonable ansatz,
however: \beq \label{e21}
f^{(1b)}_{G|\gamma} (x,Q^2,P^2) &= \Gg(x,Q^2), \hspace*{5.5cm} P^2 \leq P_c^2
\nonumber \\
&= c_G(x,Q^2) \left[ \ln \frac{Q^2}{P^2} - \ln \frac{P^2}{\Lambda^2}
\ln \frac {\ln (Q^2/\Lambda^2) }
{ \ln (P^2/\Lambda^2) } \right], \ P^2 \geq P_c^2. \eeq
In this case the continuity condition at $\psq=\pcsq$ can still easily be
solved for $c_G\xqsq$, but an explicit analytical expression for \pcsq\ for
given $c_G$ is no longer possible.

Both the ansatz (\ref{e19}) and (\ref{e21}) allow to compute the \psq\
integral in eq.(\ref{e9}) analytically. In the former case one has: \beq
\label{e22}
f^{(1a)}_{G|e}(y) = \int_y^1 \frac {dx}{x} \tilde{f} \left( \frac {y}{x}
\right) \Gg\xqsq & \left[ \ln \frac {\min (P_c^2,P_{\rm max}^2) }
{P^2_{\rm min}} \nonumber \right. \\ & \left.
+ \frac{1}{3} \theta(P_{\rm max}^2 - P_c^2) \left( \ln \frac {Q^2}{P_c^2}
- \frac { \ln^3 (Q^2/P^2_{\rm max}) } { \ln^2 (Q^2/ P_c^2) }
\right) \right]. \eeq
In a no--tag situation, $\pmaxsq=\qsq$, this simplifies to \be \label{e23}
f^{(1a), {\rm no-tag}}_{G|e}(y) = \int_y^1 \frac {dx}{x} \tilde{f} \left(
\frac {y}{x} \right) \Gg\xqsq \ln \frac {Q^2} {P^2_{\rm min}}
\left[ 1 - \frac {2}{3} \frac { \ln (Q^2/P^2_c) }
{ \ln (Q^2/ P_{\rm min}^2) } \right]. \ee
Note the similarity to the corresponding result (\ref{e13}) for \fqe. However,
the stronger suppression of the gluon density at large \psq\ leads to a
larger coefficient of the subtraction term in the square bracket (2/3 rather
than 1/2).

The somewhat more complicated ansatz (\ref{e21}) gives (for $\pmaxsq >
\pcsq$): \beq \label{e24}
f^{(1b)}_{G|e}(y) = \int_y^1 \frac {dx}{x} \tilde{f} \left( \frac {y}{x}
\right) &\left\{ \Gg\xqsq  \ln \frac {P_c^2} {P^2_{\rm min}}
\nonumber \right. \\ & \left.
+ \frac{1}{2} c_G(x,Q^2) \left[ \ln^2 \frac{Q^2}{P_c^2}
-  \ln^2 \frac{Q^2}{P_{\rm max}^2} +  \ln^2 \frac{P_c^2}{\Lambda^2}
\left( \ln \frac {\ln (Q^2/\Lambda^2) }
{ \ln (P_c^2/\Lambda^2) } + \frac {1}{2} \right)
\right. \right. \nonumber \\ & \left. \left. \hspace*{2.3cm}
- \ln^2 \frac{P_{\rm max}^2}{\Lambda^2}
 \left( \ln \frac {\ln (Q^2/\Lambda^2)}
{ \ln (P_{\rm max}^2/\Lambda^2) } + \frac {1}{2} \right)
\right] \right\}. \eeq
This expression also simplifies somewhat in the no--tag case $\pmaxsq=\qsq$,
but one does not recover a result as simple as eq.(\ref{e13}) or (\ref{e23}).
Of course, in eq.(\ref{e24}) $c_G \xqsq$ is related to \pcsq\ and \Gg\xqsq\
via the continuity condition at $\psq=\pcsq$.

As in case of the quark density, $c_G$ can be obtained from a simple
parton--level calculation. Specifically, in the picture of a photon to quark
to gluon splitting used in deriving eq.(\ref{e18}) one finds the following
$x-$dependence: \beq \label{e25}
c_G^{QPM}(x) &\propto \int_x^1 \frac{dy}{y} c_q^{QPM}(y) P_{Gq}\left(
\frac{x}{y} \right) \nonumber \\
&= N \left[\frac{4}{3} \left( \frac{1}{x} - x^2 \right) + 1 - x + 2 (1+x) \ln x
\right], \eeq
where $P_Gq$ is the quark $\rightarrow$ gluon splitting function \cite{26}.
The normalization $N$ of $c_G^{QPM}$ can be fixed from the result (\ref{e15})
for $c_q^{QPM}$, taking into account that the dependence on \psq\ and \qsq\
has already been factored out in the ansatz (\ref{e19}): \be \label{e26}
N = \frac{\alpha}{\pi} \ \frac {9} {33 - 2 N_f}  \sum_{q,\bar{q}} e_q^2, \ee
where $N_f$ is the number of active flavors. Together with the continuity
relation (\ref{e20}), eqs.(\ref{e25}) and (\ref{e26}) can again be used to
determine \pc\ for given $x$ and \qsq; of course, the result will depend on
\Gg\xqsq, which is still not very well determined experimentally \cite{14}.

Both eq.(\ref{e19}) and (\ref{e21}) suffer from the same problem as the
simple ansatz (\ref{e10}) for \fqg: The derivative with respect to $\ln \!
\psq$ is discontinuous at $\psq = \pcsq$. This can be solved in complete
analogy to eq.(\ref{e16}) by modifying the ansatz for \fGg\ to: \be \label{e27}
f^{(2)}_{G|\gamma}(x,Q^2,P^2) = \Gg \xqsq \frac { \ln^2 \frac {Q^2+P_c^2}
{P^2+P_c^2} } {\ln^2 \left( 1 + \frac {Q^2}{P_c^2} \right) }. \ee
The same procedure that led to eq.(\ref{e17}) again allows to find an
excellent approximation for the \psq\ integral in eq.(\ref{e9}):
\beq \label{e28}
f^{(2)}_{G|e}(y) = \int_y^1 & \tilde{f} \left( \frac {y}{x} \right) \Gg \xqsq
\left[ \ln \frac {P_c^2} {P^2_{\rm min}} + \frac{1}{3} \ln \left( 1 +
\frac {Q^2}{P_c^2} \right) + \frac { 2 P^2_{\rm min}/P_c^2 - \pi^2/3}
{\ln \left( 1 + \frac {Q^2}{P_c^2} \right) } \nonumber \right. \\ & \left.
+ \frac { \frac{9}{4} - \frac{1}{3} \ln^3 \frac {Q^2+P_c^2} {P^2_{\rm max}}
+ \frac {2 P_c^2}{P^2_{\rm max}} \left( \ln \frac {Q^2+P_c^2} {P^2_{\rm max}}
-1 \right) + \frac {P_c^4} {2 P^4_{\rm max}} \left( \frac{1}{2} -
\ln \frac {Q^2+P_c^2}{P^2_{\rm max}} \right) }
{ \ln^2 \left( 1 + \frac {Q^2}{P_c^2} \right) } \right] , \eeq
where we have again assumed $\pmaxsq > \pcsq$, and terms of ${\cal O}
\left( \frac{P^6_c}{P^6_{\rm max}}, \frac {P^4_{\rm min}} {P^4_c} \right)$
have been omitted. Numerically eq.(\ref{e28}) reproduces the exact result in
eq.(\ref{e9}) to better than 2\%.

\section*{5) Numerical examples}
We are now in a position to present numerical examples for \fie, using the
results of secs. 3 and 4. We start with two examples relevant for the \eplem\
collider TRISTAN, which now operates at $\rts \simeq 57$ GeV. Here we are
only interested in the reduction of the expected parton flux due to the
virtuality of the exchanged photons as well as due to experimental
(anti--)tagging conditions. We therefore normalize our results to the
most naive ``unsuppressed" prediction for \fie, which has been obtained
from eqs.(\ref{e7}) and (\ref{e5}) with $\pmaxsq = s \cdot (1-\xg)$
(eq.(\ref{e4b}) with $\theta_{\rm max} = \pi$). As already explained in sec.2
this ansatz over--estimates the correct parton flux even in the absence of
anti--tagging and high--\psq\ suppression, since the relevant scale \qsq\ of
the hard \gamgam\ scattering (to be identified, e.g., with the squared
transverse momentum of high$-p_T$ jets) is usually (much) smaller than
$P^2_{\rm{max,kin}}$.

In figs.1a (for $u-$quarks) and b (for gluons) we have chosen $\qsq=10$
GeV$^2$, typical for current \gamgam\ data at TRISTAN \cite{1,2}; no
(anti--)tagging has been required. The dotted curves show the reduction that
results from imposing the dynamical bound $\psq \leq \qsq$ on the virtuality
of the exchanged photons. These curves are only slightly affected by the
$x-$dependence of the parton densities. For very large photon energy \xg\
the kinematical constraint (\ref{e4b}) will give a bound below \qsq\ even for
$\theta_{\rm max} = \pi$ (no--tag); requiring $\psq \leq \qsq$ does therefore
not affect the flux of very energetic photons. If the parton density in the
photon is very soft (concentrated at small $x$) the region of large \xg\ will
contribute more to the convolution integral defining \fie. Therefore the
effect of requiring $\psq \leq \qsq$ is slightly smaller for the (soft) gluon
density than for the (hard) quark density.

The solid and dashed curves in figs.1 show our estimates of the combined
suppression due to the virtuality of the exchanged photons and the
requirement $\psq \leq \qsq$. The short dashed curve in fig.1a shows the
prediction (\ref{e13}) of the simple ansatz (\ref{e10}) with fixed $\pcsq =
0.3$ GeV$^2$, while the long dashed curve is the prediction (\ref{e13}) if
\pcsq\ is estimated from the QPM using eq.(\ref{e15}). The corresponding
curves in fig.1b refer to the prediction (\ref{e23}) of the simple ansatz
(\ref{e19}) with fixed \pcsq, and with \pcsq\ determined from the QPM
results (\ref{e25}), (\ref{e26}), respectively; the dot--dashed curve here
shows the prediction (\ref{e24}) of the somewhat more complicated ansatz
(\ref{e21}). In these figures the solid curve shows predictions
(eqs.(\ref{e17}), (\ref{e28})) of the smoothed--out ans\"atze (\ref{e16})
and (\ref{e27}), respectively, where we have assumed $\pcsq=0.5$ GeV$^2$.

We see that all predictions are quite similar. Notice, however, that we have
used slightly different values for \pcsq\ with the smooth ans\"atze for \fig\
than for the simple ones whose derivatives are discontinuous. This is
reasonable since the former predict some suppression for all $\psq \ne 0$,
while the latter assume \fig\ to be completely unsuppressed for $\psq <
\pcsq$. The fact that the results using the QPM estimates (\ref{e15}) and
(\ref{e25}), (\ref{e26}) come out quite close to the other predictions gives
us some confidence that our choices of \pcsq\ are indeed
reasonable.\footnote{When using the QPM estimates we have always required
$\pcsq \geq \lsq$, see eq.(\ref{e14}), i.e. we have set $\pcsq=\lsq (=0.04$
GeV$^2$ for the GRV parametrization \cite{9} used in fig.1) if the QPM
predicts $\pcsq < \lsq$. This happens only at small $x$, where (multiple)
gluon radiation is expected to be important, so that the QPM prediction
cannot be trusted.} We should mention, however, that this result depends to
some extent on the parametrization of the parton densities in the photon. For
example, when used with the DG parametrization \cite{7} the QPM predicts
significantly less suppression of $f_{u|e}$ at large $x$. The reason is that
this parametrization has a rather small $u^{\gamma}$ at large $x$, which
implies a large value of \pcsq, and hence little suppression, if $c_u$ is
fixed from the QPM, see eq.(\ref{e11}). Generally we conclude that in a
no--tag situation with $\qsq=10$ GeV$^2$ virtual photon effects suppress
\fqe\ by 8 to 10\% and \fGe\ by 12 to 15\% even after the constraint $\psq
\leq \qsq$ has been included; one expects even larger suppression at larger
\qsq, since then a larger fraction of the \psq\ integral in eq.(\ref{e9})
comes from the region $\psq > \pcsq$ where the \fig\ are reduced
significantly.\footnote{In order to avoid confusion we should mention that
the absolute values of the \fie\ still increase with increasing \qsq\ even
after the suppression of virtual photon structure functions has been taken
into account. However, the increase is slower than one would expect in the
absence of this suppression; therefore in a no--tag situation the suppression
becomes relatively more important at larger \qsq.} Apart from the region of
large $x$, which contributes only little to any cross section because $\fie
\rightarrow 0$ as $x \rightarrow 1$, the predicted suppression is almost
independent of $x$ if a fixed value of \pcsq\ is assumed; this is not true if
\pcsq\ is estimated from the QPM, however.

In fig. 2a,b we show corresponding results for an anti--tag situation; this
might be more relevant for practical applications, since some (anti--)tagging
is usually applied in experimental analyses \cite{1,2,3} of two--photon data,
in order to separate events with low and high \psq. In these figures we have
used the anti--tagging applied by the TOPAZ collaboration in their recent
analysis of jet production in \gamgam\ collisions: $\theta_{\rm max} =
3.2^{\circ}$ for scaled photon energy $\xg \leq 0.75$, and $\theta_{\rm max}
=\pi$ otherwise. For easier comparison with fig.1 we use the same
``unconstrained" \fie\ as before, where neither anti--tagging nor the bound
$\psq \leq \qsq$ has been taken into account. Both these constraints have been
included in the dotted curves in fig.2, which (for $\xg < 0.75$) therefore lie
significantly below the corresponding curves in fig.1 where no anti--tagging
was assumed. Notice, however, that anti--tagging has lowered our final result
for the \fie\ (solid and dashed curves), including virtual photon effects, by
only 2--3\% compared to fig.1. The reason is that the region $\psq > P^2_{\rm
max,tag}$ contributes relatively little to \fie\ even in fig.1, due to the
suppression of the \fig\ at these high \psq. Once anti--tagging has been taken
into account, virtual photon effects suppress \fqe\ by only about 5\%. In case
of gluons, however, this additional reduction could be as large as 10\% even
in the region of small $x$ where \fGe\ is sizable. Our anti--tagging condition
is less effective for gluons since, as discussed above, \fGe\ gets a
relatively larger contribution from the region of large \xg\ than \fqe\ does.
This demonstrates that the exact experimental implementation of anti--tagging
is important. In the present case only electrons with energy $> E_{\rm
beam}/4$ are vetoed at large angles, which does not affect events where most
of the energy of the incident electrons is carried away by the photon.

Notice also that the QPM prediction for \fGe\ (long dashed curve in fig. 2b)
now differs significantly from the predictions for fixed \pcsq\, at least
in the region $x > 0.4$; this is because we have used the DG parametrization
\cite{7} here. However, this region will not contribute much to any cross
section, since this parametrization is characterized by a rather soft gluon
distribution function. For practical purposes our different ans\"atze for
\fig\ therefore still give quite similar results. In particular,
$f^{(1b)}_{G|e}$ of eq.(\ref{e24}) always comes out very close to
$f^{(1a)}_{G|e}$ of eqs.(\ref{e22}) and (\ref{e23}); the use of the somewhat
more cumbersome ansatz (\ref{e21}) for \fGg\ therefore hardly seems worth the
trouble, considering that it still suffers from a discontinuous derivative
at $\psq=\pcsq$.

Our final example concerns the small angle electron tagger that has recently
been installed \cite{27} in the ZEUS detector at the $ep$ collider HERA.
Unlike the forward taggers used by both HERA experiments \cite{4,5}, this
detector is only sensitive to events with a finite, although small, photon
virtuality: 0.1 GeV$^2 \leq \psq \leq$ 1 GeV$^2$. At HERA ($\rts \simeq 296$
GeV at present) this implies $P^2_{\rm min,tag} = 0.1 \ {\rm GeV}^2 >
P^2_{\rm min,kin}$ and $P^2_{\rm max,tag} = 1.0 \ {\rm GeV}^2 < P^2_{\rm
max,kin}$ for almost all photon energies. The predicted suppression of the
\fie\ due to virtual photon effects is therefore independent of $x$ if one
of our ans\"atze is used with fixed \pcsq.

The suppression does depend on the scale \qsq\ characterizing $\gamma p$
scattering, however. This is demonstrated in fig.3, where we show the
suppression of $f_{u|e}$ (solid) and \fGe\ (dashed) for $\pcsq=0.15$ and
0.5 GeV$^2$, as predicted from the simple ans\"atze (\ref{e10}) and
(\ref{e19}). In contrast to the situation depicted in fig.1 the suppression
now decreases with increasing \qsq. The reason is that here, unlike in fig.1,
the upper limit of the \psq\ integration in eq.(\ref{e9}) is simply fixed by
the experimental tagging condition, which is independent of \qsq. The
behaviour depicted in fig.3 then follows from the fact that $\fig(\psq > \pcsq
)$ is relatively less suppressed at larger \qsq. Experimentally \qsq\ can,
e.g., be identified with the squared transverse momentum of high--$p_T$ jets
produced in the event. The ratio shown in fig.3 can therefore be measured
experimentally by comparing the rate for jet events where the electron is
detected in the small angle tagger to that where the electron hits the
forward spectrometer presently used for tagging photoproduction events (and
for measuring the luminosity); this forward spectrometer only accepts events
with $\psq < 0.1 \ {\rm GeV}^2$, where the virtuality of the photon should
indeed be negligible.\footnote{The effective photon flux for events tagged by
the forward spectrometer is considerably larger than for events tagged by the
small angle tagger; in the former case, $\ln ( \pmaxsq / \pminsq ) \simeq
13$, compared to 2.3 for the latter. One can easily correct for this known
difference in photon fluxes to determine the suppression due to the virtuality
of the photon, shown in fig.3.} This measurement should be very clean since by
comparing events with equal characteristics of the hadronic system ($p_T$ and
rapidity of the jets) and with equal energy of the tagged electron most
hadronic uncertainties, e.g. related to unknown structure functions, will
cancel out.

As usual we find larger suppression for the gluon density than for quark
densities. Notice that on average gluon--induced jet events look quite
different from quark--induced and direct events \cite{28}: The high--$p_T$
jets tend to emerge at larger rapidities, closer to the proton beam direction;
and they tend to have more energetic photon remnant jets. Since going to
finite \psq\ suppresses gluon--induced processes more than quark--induced
processes while direct processes are not suppressed at all (apart from the
trivial reduction of \fge) we expect the high--$p_T$ jets in events tagged by
the small angle tagger to be on average more central, compared to events
tagged by the forward spectrometer; similarly, the former class of events
should on average have somewhat less hadronic activity from the photon
remnants in the electron beam direction. These qualitative effects can
unambiguously be predicted from QCD, which requires \fGe\ to be more
strongly suppressed than \fqe\ \cite{19}; however, fig.3 shows that the size
of these effects depends on the nonperturbative parameter \pcsq, which at
present cannot be predicted from first principles. We should mention here
that we regard the values of \pcsq\ chosen in fig.3 to approximate the lower
and upper bound of the range of reasonable values; our ``best guess",
corresponding to the value chosen in figs. 1 and 2, would fall roughly halfway
in between these two.

\section*{6) Summary and Conclusions}
In this paper we have studied the reduction of the effective parton flux in
the electron due to experimental (anti--)tagging, as well as due to the
suppression of virtual photon structure functions compared to the more
familiar structure functions of real (on--shell) photons. Our main results are
given in eqs.(\ref{e12}) and (\ref{e17}) for quark densities and
eqs.(\ref{e22}), (\ref{e24}) and (\ref{e28}) for gluon densities. These
effects treat the dependence of photon structure functions on the virtuality
\psq\ of the photon only in an approximate manner; however, we argued in sec.3
that they ought to reproduce the relevant integrals over \psq\ quite
accurately, since they are based on parametrizations of the parton content of
virtual photons that have the correct low and high \psq\ limits. The virtue of
this simplified approach is that it still allows to express the effective
parton densities in the electron in terms of a single convolution integral,
similar to the standard expression (\ref{e7}) where the reduction of virtual
photon structure functions is ignored. These parton densities in the electron
directly enter predictions for cross sections of resolved photon processes, as
shown in eq.(\ref{e8}).

In our numerical examples of sec.5 we found that the size of the suppression
depends both on the experimental (anti--)tagging requirements and on the
scale \qsq\ at which the photon is probed. In a no--tag situation \qsq\
provides the upper limit on \psq, since for $\psq > \qsq$ it no longer makes
sense to describe the process in terms of partons ``in" the (virtual) photon.
In this situation increasing \qsq\ gives more relative weight to the region
of large (compared to \lsq) photon virtualities, and hence leads to larger
suppression factors. Conversely, if experimental (anti--)tagging determines
the upper bound on \psq, increasing \qsq\ will reduce the suppression factor
since the relative (to \qsq) virtuality of photons in accepted events is
reduced.

Present \gamgam\ experiments are now analyzing data with \qsq\ typically
around 10 GeV$^2$. We estimate that in a no--tag situation virtual photon
effects then suppress the effective quark and gluon content by about 10 and
15\%, respectively; note that the reduction of cross sections of
twice--resolved \gamgam\ processes is twice as large, since they contain two
factors of \fie. Under experimentally more relevant anti--tagging conditions
we estimate the suppression of quark densities to be a modest 2--3\%, which is
hardly significant compared to other experimental and theoretical
uncertainties; however, gluon densities could still be reduced by 10\%, an
effect similar in size to the recently computed NLO corrections to jet
production in real \gamgam\ scattering \cite{24}.

We finally pointed out that the small angle electron tagger recently
installed in the ZEUS experiment at HERA should allow to study the onset of
the suppression of virtual photon structure functions in some detail. Hadronic
uncertainties can largely be removed by comparing the rate of events tagged
by this device to that tagged by the existing forward spectrometer. Our
``best guess" for the suppression at $\qsq=100$ GeV$^2$ is around 8 and 15\%
for quark-- and gluon--initiated processes, respectively. The stronger
suppression of the rate of events with a gluon from the photon in the initial
state should lead to changes in the average rapidity of the hard jets as well
as the average energy of the photon remnant. We remind the reader here that we
ignored the possible existence of a ``soft" contribution to the photon
structure function. Such a contribution would be much more strongly suppressed
at high \psq, and should therefore be easily detectable by this small angle
tagger.

In summary, effects due to the suppression of virtual photon structure
functions are of roughly comparable size as NLO QCD corrections in a no--tag
experiment; they are somewhat smaller, but can still be non--negligible, when
anti--tagging is imposed. They should therefore be taken into account when
one tries to extract the parton densities in real photons from \gamgam\ data
taken at \eplem\ colliders. An experiment that allows to tag outgoing
electrons both in the forward direction and at small but nonvanishing angles
has the opportunity to study these effects in some detail, thereby shedding
new light on the interplay between soft and hard QCD.

\subsection*{Acknowledgements}
The work of M.D. was supported in part by the U.S. Department of Energy under
contract No. DE-AC02-76ER00881, by the Wisconsin Research Committee with funds
granted by the Wisconsin Alumni Research Foundation, by the Texas National
Research Laboratory Commission under grant RGFY93--221, as well as by a grant
from the Deutsche Forschungsgemeinschaft under the Heisenberg program.

\clearpage
\section*{Figure Captions}
\renewcommand{\labelenumi}{Fig.\arabic{enumi}}
\begin{enumerate}

\item    
The reduction of the parton density in the electron in a no--tag situation due
to the bound $\psq \leq \qsq$ as well as due to the suppression of virtual
photon structure functions. All curves are normalized to the parton densities
obtained from eqs.(\ref{e4})--(\ref{e7}) with $\theta_{\rm max} = \pi$,
ignoring the condition (\ref{e6b}); a) is for the $u-$quark density, while b)
is for the gluon density. The dotted curves show the effect of only requiring
$\psq \leq \qsq$, while the dashed and solid curves also include the
suppression of \fig\ at $\psq \ne 0$. In a), the short dashed and long dashed
curves represent the prediction (\ref{e13}) with fixed \pcsq\ and with \pcsq\
estimated from the QPM, respectively, while the solid line shows the result
(\ref{e17}). In b), the short dashed and long dashed curves depict the
prediction (\ref{e23}) with fixed \pcsq\ and with \pcsq\ estimated from the
QPM, respectively, while the dot--dashed and solid lines represent the
predictions from eqs.(\ref{e24}) and (\ref{e28}), respectively. The leading
order parametrization of ref.\cite{9} has been used.

\vspace*{5mm}
\item   
The reduction of $f_{u|e}$ (a) and \fGe\ (b) for an anti--tag situation, i.e.
eq.(\ref{e4b}) has been used with $\theta_{\rm max}=3.2^{\circ}$ iff $\xg \leq
0.75$. We have used the parametrization of ref.\cite{7}. Notations are as in
fig.1.

\vspace*{5mm}
\item  
The reduction of the parton flux in the electron due to the suppression of
$\fig(\psq \ne 0)$ if the photon virtuality is restricted to lie in the
range 0.1 GeV$^2 \leq \psq \leq$ 1.0 GeV$^2$, predicted from eqs.(\ref{e12})
(for $f_{u|e}$, solid curves) and (\ref{e22}) (for \fGe, dashed) with fixed
\pcsq. Since the limits on \psq\ are independent of $x$ the reduction of the
parton fluxes also does not depend on $x$, nor on the parametrization of real
photon structure functions chosen. It does, however, depend on the scale \qsq\
at which the photon is being probed, as shown in the figure. This suppression
should be measurable at the ZEUS detector at HERA, as discussed in the text.

\end{enumerate}
\end{document}